# Absence of metallicity and bias-dependent resistivity in low-carrier-density EuCd$_2$As$_2$

Yuxiang Wang[1], Jianwen Ma[1], Jian Yuan[2], Wenbin Wu[3], Yong Zhang[4], Yicheng Mou[1], Jiaming Gu[1], Peihong Cheng[2], Wu Shi[1,6], Xiang Yuan[3,5], Jinglei Zhang[4*], Yanfeng Guo[2,7*], Cheng Zhang[1,6*]

[1] State Key Laboratory of Surface Physics and Institute for Nanoelectronic Devices and Quantum Computing, Fudan University, Shanghai 200433, China
[2] School of Physical Science and Technology, ShanghaiTech University, Shanghai 201210, China
[3] State Key Laboratory of Precision Spectroscopy, East China Normal University, Shanghai 200241, China
[4] Anhui Province Key Laboratory of Condensed Matter Physics at Extreme Conditions, High Magnetic Field Laboratory of the Chinese Academy of Sciences, Hefei 230031, China
[5] School of Physics and Electronic Science, East China Normal University, Shanghai 200241, China
[6] Zhangjiang Fudan International Innovation Center, Fudan University, Shanghai 201210, China
[7] ShanghaiTech Laboratory for Topological Physics, Shanghai 201210, China

* Correspondence and requests for materials should be addressed to J. Z. (E-mail: zhangjinglei@hmfl.ac.cn), Y. G. (E-mail: guoyf@shanghaitech.edu.cn) & C. Z. (E-mail: zhangcheng@fudan.edu.cn)

**Abstract**

EuCd$_2$As$_2$ was theoretically predicted to be a minimal model of Weyl semimetals with a single pair of Weyl points in the ferromagnet state. However, the heavily *p*-doped EuCd$_2$As$_2$ crystals in previous experiments prevent direct identification of the semimetal hypothesis. Here we present a comprehensive magneto-transport study of high-quality EuCd$_2$As$_2$ crystals with ultralow bulk carrier density ($10^{13}$ cm$^{-3}$). In contrast to the general expectation of a Weyl semimetal phase, EuCd$_2$As$_2$ shows insulating behavior in both antiferromagnetic and ferromagnetic states as well as surface-dominated conduction from band bending. Moreover, the application of a *dc* bias current can dramatically modulate the resistance by over one order of magnitude, and induce a periodic resistance oscillation due to the geometric resonance. Such nonlinear transport results from the highly nonequilibrium state induced by electrical field near the band edge. Our results suggest an insulating phase in EuCd$_2$As$_2$ and put a strong constraint on the underlying mechanism of anomalous transport properties in this system.

## 1. Introduction

Topological phases of matter have attracted considerable interest in the past two decades owing to the topological nontrivial band structure and gapless surface state[1–5]. Thereinto, due to the coupling of electronic topology and spin configuration, magnetic topological materials have been intensively investigated towards the realization of dissipationless chiral charge and spin transport[6]. Extensive progress has been made in the research of magnetic topological insulators, which leads to the discovery of quantum anomalous Hall effect[7,8]. Most recently, magnetic Weyl semimetals (WSM) stand out as a promising candidate for exploring the combination of topology and magnetism in metallic systems. However, magnetic topological materials typically show complicated electronic structure, which further evolves with external magnetic fields[9–15]. Therefore, the realization of a simple magnetic WSM material would be conducive to further experimental exploration.

A minimal model of magnetic WSM consists of a single pair of Weyl points (WPs) splitted by the breaking of time reversal symmetry. Such a simple magnetic WSM system could largely simplify the research of Weyl physics, such as chiral anomaly, anomalous Hall effect as well as Weyl orbit[16–21]. Among various proposals[5,22–24], $EuCd_2As_2$ was recognized as a promising candidate of magnetic WSM with a single pair of WPs in ferromagnetic (FM) phase[16,25–41]. $EuCd_2As_2$ is an A-type antiferromagnet (AFM) below Néel temperature $T_N$ = 9.5 K with space group 164 ($P\bar{3}m1$)[28]. It was predicted to be an AFM Dirac semimetal (an AFM-topological insulator) with (without) the protection of $C_{3z}$ symmetry[29]. The Eu spins align parallel to the triangular layers and break the $C_{3z}$ symmetry as revealed by resonant elastic x-ray scattering[28,39]. When the Eu spins are switched to the $c$ direction (perpendicular to the triangular layers) by magnetic fields, a pair of WPs will be generated by the band crossing induced by FM exchange interaction[16,28]. It results in a topological phase transition from AFM-topological insulator to FM-WSM. The WPs can be also modulated by spin-canting in magnetic fields[30,32,40]. Various transport phenomena including negative magnetoresistance (MR), anomalous transverse transport[25–28,33] were explained in such scenario. Nevertheless, previous experimental works[25,26,28,32] mainly focused on the heavily $p$-doped $EuCd_2As_2$ crystals, which prevents a direct identification of the semimetal hypothesis. It remains elusive whether a gapless magnetic WSM is indeed generated in the field-induced FM state.

Here we report a comprehensive magneto-transport study of high-quality $EuCd_2As_2$ with ultralow bulk carrier density ($10^{13}$ cm$^{-3}$). In contrast to the metallic behavior in heavily $p$-doped samples, these low-carrier-density $EuCd_2As_2$ crystals show much larger resistivity with insulating-like temperature dependence in both AFM and FM states, while featuring similar magnetization properties. In these semiconducting samples, we find that the surface conduction dominates the transport due to band bending, which is removed after mechanical polish. Meanwhile, the resistance can be effectively tuned by the $dc$ bias current due to Fermi surface displacement in electric fields. The dramatic decrease of resistance points to enhanced mobility at large bias, which indicates a curved band dispersion rather than a linear one. In such a highly nonequilibrium state, Hall-field induced resistance oscillations (HIRO) can be detected as functions of magnetic fields and bias current. These complex transport results in $EuCd_2As_2$ call for further investigations on the low-energy electronic structure of this system.

## 2. Results
### 2.1 Magneto-transport and surface-dominated conduction

Electrical transport was conducted in the (001) plane of the pristine as well as polished $EuCd_2As_2$ crystals with current along the $a$ axis and magnetic field along the $c$ axis. The $EuCd_2As_2$ crystals were synthesized by using Tin as flux. Polished samples were polished by sandpapers with a home-made handheld tool. The resistivity-temperature ($\rho$-$T$) curve was plotted in Fig. 1(a). Surprisingly, for the same piece of sample, the resistivity at 300 K increases by over an order of magnitude after surface mechanical polish compared to the pristine crystal before polish. The resistance difference further increases up to over three orders of magnitude at 150 K. The main difference between pristine and polished crystals is that the initial surface layer was replaced by a polished sublayer with enhanced roughness (Fig. S1(d)). The dramatic change in resistivity suggests that the surface layer accounts for over 90% of the total conduction at 300 K, which is further enhanced at low temperatures. To explore the origin of such anomaly, we measured the magnetic field $B$ dependence of $R_{xx}$ and $R_{yx}$ as shown in Fig. 1 (b-c). Hall coefficient of the polished sample is $-8\times10^{-2}$ m$^3\cdot$C$^{-1}$ at 300 K, which is also an order of magnitude larger than that of the pristine crystal before polish ($-2.8\times10^{-3}$ m$^3\cdot$C$^{-1}$) as shown in Fig. 1(b). As temperature reduces, the polished sample presents $n$-type Hall coefficient with thermal activation behavior down to 150 K with an activation energy of $\Delta_A = 213.3$ meV (Fig. 1(a), inset). Different from the polished one, the pristine sample presents a sign change of Hall coefficient from minus ($n$-type) to plus ($p$-type) at around 270 K along with a slightly smaller activation energy of $\Delta_B = 137.5$ meV. The temperature dependence of Hall coefficients before and after polishing is summarized in Fig. 1(d). It corresponds to an insulating bulk state in the polished sample with carrier density down to $2.8\times10^{13}$ cm$^{-3}$ at 210 K, below which the Hall coefficient is hard to obtain due to the exceptionally large resistance. Note that the carrier density here is much smaller than the value in previous experiments[25,26,28] on heavily $p$-doped EuCd$_2$As$_2$ ($10^{19}$ cm$^{-3}$) despite similar magnetization properties (Fig. S3). In contrast, the pristine crystals in our experiment present comparably less insulating behavior due to the presence of additional hole carriers at the surface. From the Hall coefficient, we roughly estimate the surface hole density as $5.3\times10^{13}$ cm$^{-2}$ at 210 K. The fitting of Hall conductivity based on the two-carrier model[42] is shown in Fig. S9 and Fig. S10.

To investigate the nature of the surface layer, we performed systematical magneto-transport and magnetization measurements in the pristine crystals down to 2 K (Fig. 2). Note that here and after, the electric transport parameters for the pristine crystals are presented in the form of two-dimensional systems due to the surface-dominating conduction. The temperature dependence of longitudinal resistance $R_{xx}$ (Fig. 2(a), inset) show two distinct regions (2-100 K and 270-300 K, respectively), where resistance increases exceptionally with the decrease of temperature and reaches a very high value of 4.3 MΩ at 2 K. The Fig. S2(a) highlights a kink in the $R_{xx}$-$T$ curve near the AFM transition. In Fig. 2(a) and 2(b), we show that negative magnetoresistance and anomalous Hall effect occur due to the AFM-to-FM transition in magnetic fields with spin aligned to the $c$ axis. Such behavior is similar to heavily $p$-doped EuCd$_2$As$_2$ but with a much larger ratio. $R_{xx}$ decreases for nearly two orders of magnitude from 4.3 MΩ at 0 T to 0.089 MΩ at 2 T at 2 K[26,32]. Hall resistance data between -1 T and 1 T below 10 K was removed due to the impact of rapid decrease in $R_{xx}$, which cannot be well removed by the anti-symmetrization process. Fig. 2(c) presents the field dependence of magnetization with consistent saturation field with MR in Fig. 2(a). The temperature dependence of magnetization (Fig. 2(c), inset) is consistent with previous studies[25,28], which gives the Néel temperature $T_N = 9.5$ K. The carrier density and in-field mobility in the FM state are calculated as presented in Fig. 2(d). Note that due to the band shift induced by magnetic fields, the carrier density calculated from Hall effect here only corresponds to the FM state in the magnetic field rather than the zero-field AFM state. We find that Hall carrier density $n_H$ shows clear decrease from 20 K to 2 K while mobility at 3 T $\mu(3\text{T})$ stays similar. From the decrease of $n_H$ towards low $T$ and the anomalous large resistance (Fig. S2(b)), it is reasonable to consider an insulating phase for the field-enforced FM state in EuCd$_2$As$_2$ that was presumed to be a Weyl semimetal in theory.

**2.2 Nonlinear transport**

To provide further evidence for the above argument, we performed nonlinear transport experiment in the pristine crystals as illustrated in Fig. 3(a). Differential resistance $R_{xx}^{\text{Diff}}$ and $R_{yx}^{\text{Diff}}$ (Fig. 3(b)) was measured using a small *ac* current superimposed on a *dc* bias current $I_{dc}$ while monitoring the *ac* voltage. Both the AFM and FM states can be effectively tuned by the application of relatively small $I_{dc}$ of several µA. Differential resistance at zero magnetic field $R(0\text{T})$ changes from 4.3 MΩ to 1.8 MΩ at 3 µA, while negative MR ratio $R(0\text{T})/R(3\text{T})$ was tuned from 51 to 13. Fig. 3(c) is the complete $R_{xx}^{\text{Diff}}$-$I_{dc}$-$B$ diagram, showing a cross-like high-resistance-state region emerging along the zero-magnetic-field and zero-bias-current lines, respectively. By linear fitting of $R_{yx}^{\text{Diff}}$, we find that the carrier density $n_H$ only increases slightly from $1.8\times10^{12}$ cm$^{-2}$ to $2.0\times10^{12}$ cm$^{-2}$ at 3 µA, while $\mu(3\text{T})$ shows comparably large increase from $41\,\text{cm}^2\text{V}^{-1}\text{S}^{-1}$ to $56\,\text{cm}^2\text{V}^{-1}\text{S}^{-1}$ (Fig. 3(d)). It suggests that the resistance decrease at large $I_{dc}$ mainly results from enhanced mobility. Such behavior contrasts with conventional systems where electron drift velocity tends to saturate at high electric fields[43]. Meanwhile, the bias-dependent resistance as well as the *I-V* curve (Fig. S2(c)) also differ from insulator-to-metal transitions induced by large current, which typically appears above certain threshold electric field[44]. And the weak dependence of Hall coefficients on $I_{dc}$ also excludes the electrical activation of additional mobile carriers as the origin of resistance change.

**2.3 Hall-field induced resistance oscillations**

More interestingly, oscillation patterns were detected in the low-resistance FM state. Fig. 4(a) shows typical oscillations in differential resistance versus $B$ at 10 K. The resistance oscillations only appear when large current bias is applied. They are periodic with $1/B$ as confirmed by the second derivative of $R_{xx}^{\text{Diff}}$ versus $1/B$ in the inset of Fig. 4(a) and Landau fan diagram in Fig. S7(a). And the periodicity in $1/B$ decreases with the increase of $I_{dc}$ (Fig. 4(b)) and evolves drastically with temperature (Fig. S7(c)). We also track the oscillations as a function of $I_{dc}$ as shown in Fig. 4(c-d). Prominent oscillations can be observed above 1 µA and persist up to 10 µA. Similarly, the resistance oscillations are also periodic in $I_{dc}$ and the periodicity increases with $B$. Fig. 4(a-d) together indicate that both $I_{dc}$ and $B$ are essential to the emergence of resistance oscillations. These features clearly exclude conventional quantum oscillation mechanisms such as Shubnikov-de Haas oscillation and Aharonov–Bohm oscillation, in which the oscillation periodicity is independent of bias current.

Here we introduce the HIRO scenario[45,46] that naturally fits these features. As illustrated in Fig. 4(e), let's consider a series of Landau levels tilted by transverse Hall field $E_y$ along y direction. The spatial separation between the guiding center of two tilted cyclotron orbit with $\Delta N = 1$ is $\Delta y = \dfrac{\hbar \omega_c}{eE_y}$, where $\omega_c$ is the cyclotron frequency, $N$ is the Landau level index, $e$ is elementary charge, $\hbar$ is reduced Planck's constant. HIRO results from geometric resonance in electron transitions between tilted Landau levels when the diameter of the cyclotron orbit $2R_c$ become commensurable with $\Delta y$. A geometry ratio can then be defined as $\partial_{dc} = \dfrac{2R_c}{\Delta y} = F_N \dfrac{I_{dc}}{B}$ with $F_N = \dfrac{2m^{*2}v_F}{we^2\hbar n_H}$, where $m^*$ is effective mass, $v_F$ is Fermi velocity. A more general coefficient of $F_N^{'} = w \cdot F_N$ could be used to exclude the influence of size variation, where $w$ is sample width. It then leads to periodic oscillations in both $I_{dc}$ and $1/B$ as has been observed in two dimensional electron gas and double quantum wells systems[46,47]. Following the HIRO scenario, we determine the geometry ratios by the Landau fan diagram (Fig. S7(a) and S7(b)) and label them in Fig. 4(a) and 4(c). The HIRO amplitude in EuCd$_2$As$_2$ reached a maximum at a geometry ratio of 4~6 both in sweeping $B$ and $I_{dc}$. The normalized periodicity $F_N = F_B / I_{dc}$ in $1/B$ is plotted in the left panel of Fig. 4(f). $F_N$ stays nearly unchanged with an averaged value of $8.4\ \text{T} \cdot \mu\text{A}^{-1}$ below 3 μA, corresponding to $F_N^{'} = 2.3\ \text{T} \cdot \text{mm} \cdot \mu\text{A}^{-1}$. When $I_{dc}$ exceeds 3μA, $F_N$ starts to decrease, which may result from the increased carrier density at large biases due to heating effect. The right panel of Fig. 4(f) plots $F_N = F_I \cdot B$ at different magnetic fields with $F_I$ denoting the oscillation periodicity in $I_{dc}$. The $F_N$ value from sweepings $I_{dc}$ is close to the one from sweepings $1/B$ but shows systematical increase with $B$. A more detailed investigation of the field-dependence of oscillation frequency is shown in Fig. S11. The increase of $F_N$ with $B$ may result from the additional contribution of anomalous Hall voltage.

3. Discussion

Having presented the transport results in the pristine and polished EuCd$_2$As$_2$, we now turn to discuss the implications of understanding the electronic states in this material. Different from the previous heavily p-doped samples[25,26,28], the EuCd$_2$As$_2$ crystals used in this work show much smaller carrier density, indicating less band filling and a slightly higher Fermi energy (closer to band edge). The as-grown crystals have consistent crystalline structure and magnetization properties with previous heavily *p*-doped ones. Transport results in the polished samples suggest a thermal activation behavior in the bulk of EuCd$_2$As$_2$ crystals with resistivity around $10^6$ Ω·cm at 150 K (Fig. S5(a)), eight orders of magnitude larger than that of heavily *p*-doped EuCd$_2$As$_2$[25]. By applying a large bias voltage up to 10 V, we managed to measure one of the polished samples down to 2 K (Fig. S6). The extraordinarily large resistivity and low carrier density persist in both AFM and FM states, which seems to be inconsistent with the topological semimetal hypothesis. Further investigation is required to clarify whether such insulating behavior is a result of gapped band structure or certain electron localization mechanism.

As for the dominant surface conduction in the pristine crystals, we argue that it is likely to originate from the band bending near the crystal surface (Fig. S8), which is indicated by similar mobility value of polished and pristine sample in the two-carrier-model fitting (Fig. S9 and Fig.S10). MR curves at low temperatures share similar *B*-dependence with that of heavily *p*-doped samples[26,28] but with even larger negative MR ratio. And the saturation field of negative MR is consistent with that of bulk magnetization curve. Hence it should not come from direct conduction from trivial surface contamination such as Sn flux residue, which shows distinct field dependence. We also checked the energy dispersive X-Ray spectroscopy of the pristine crystal and found no evidence of surface Sn residue (Fig. S1(b)). The nonlinear transport with non-equilibrium electron occupation provides a test for the band dispersion near the Fermi level. For a parabolic-like dispersion, Fermi velocity increases dramatically with energy near the band edge, which can lead to an increase of mobility, while staying fixed for a linearly-dispersed Dirac/Weyl cone (Fig. 3(a)). The dramatical enhancement of mobility and conductivity in electric fields (Fig. 3(d)) then points to a parabolic-like rather than a linear dispersion, which is in stark contrast to the case in graphene[48]. On the other hand, the presence of Landau levels evidenced by HIRO suggests highly-coherent transport with surface Fermi energy crossing the band rather than locating in a gap. Together these observations tend to fit with the band bending scenario[49] commonly observed in semiconductors induced by certain lattice termination (intrinsic surface state) or possible surface doping (extrinsic surface state). A recent experiment reported that surface absorption can strongly shift the Fermi energy in EuCd$_2$As$_2$ even in a high-vacuum condition[23]. Therefore, it is possible that unintentional surface doping was introduced during the growth and led to a Fermi energy shift at surface.

## 4. Conclusion

Overall, we have conducted magneto-transport in low-carrier-density EuCd$_2$As$_2$ crystals. Different from the metallic behavior in heavily *p*-doped ones, they present no sign of metallicity in both AFM and FM states. We find that the conduction in pristine crystals is mainly dominated by the surface channel and can be further tuned to highly nonequilibrium states by bias current. These complex transport behaviors in EuCd$_2$As$_2$ cannot be easily explained in the context of the proposed magnetic WSM phase and call for further experimental and theoretical investigations.

During the preparation of this manuscript, we find a similar work posted online also reporting the insulating property of intrinsic EuCd$_2$As$_2$ crystals[50].

**Acknowledgments**

We thank Xiangyu Cao and Pengliang Leng for stimulating discussions. C.Z. was sponsored by the National Key R&D Program of China (Grant No. 2022YFA1405700), the National Natural Science Foundation of China (Grant No. 12174069), and Shuguang Program from the Shanghai Education Development Foundation. Y.F.G. acknowledges the support by the Double First-Class Initiative Fund of ShanghaiTech University.

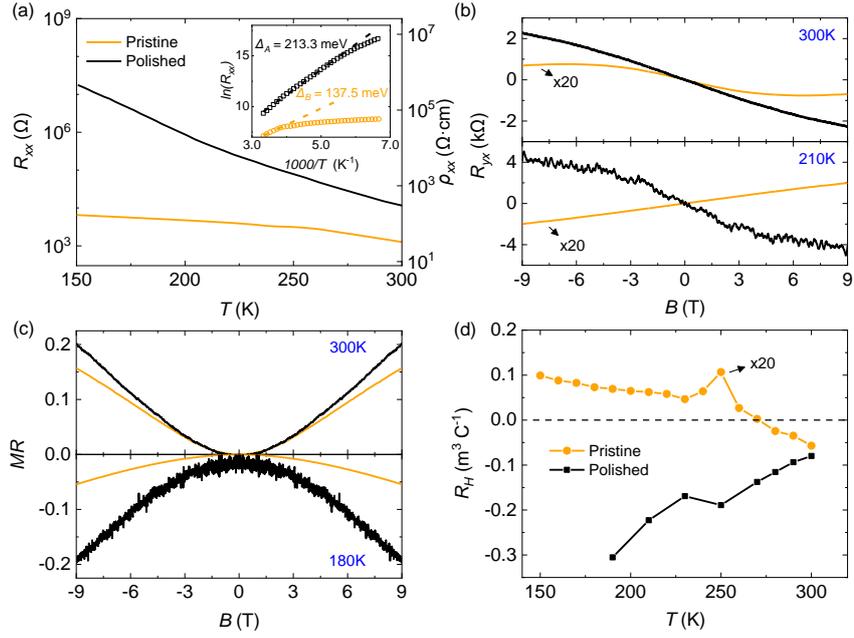

**FIG. 1. Comparison of transport properties between pristine (yellow) and polished (black) EuCd$_2$As$_2$.** (a) Sheet resistance $R_{xx}$ plotted as a function of temperature (sample A1). The right axis corresponds to calculated resistivity $\rho_{xx}$. The inset is Arrhenius plot of $R_{xx}$ and corresponding activation energies. (b-c) Hall resistance $R_{yx}$ and MR ratio at 300 K and 210 K. $R_{yx}$ for pristine sample is multiplied by 20. (d) Hall coefficient $R_H$ at different temperatures (derived from the linear fitting of Hall resistivity from -2 T to 2 T). $R_H$ for the pristine sample is multiplied by 20.

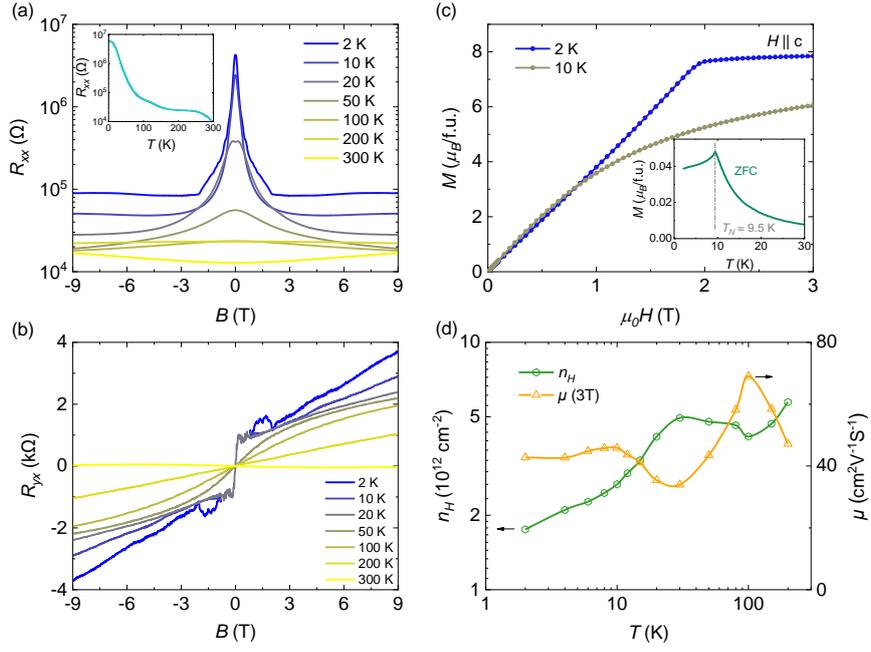

**FIG. 2. Transport and magnetization data of pristine EuCd$_2$As$_2$.** (a-b) The magnetic field dependence of $R_{xx}$ and $R_{yx}$ at different temperatures (sample A2). The inset of (a) is $R_{xx}$ plotted as a function of temperature. (c) Field sweeps of the magnetization with magnetic fields applied along *c* axis. The inset is zero-field-cooled (ZFC) magnetization data. (d) 2D carrier density $n_H$ (green hexagon) and mobility at 3 T $\mu(3T)$ (yellow triangle). $n_H$ is calculated from linear fitting of $R_{yx}$ from 3 T to 9 T.

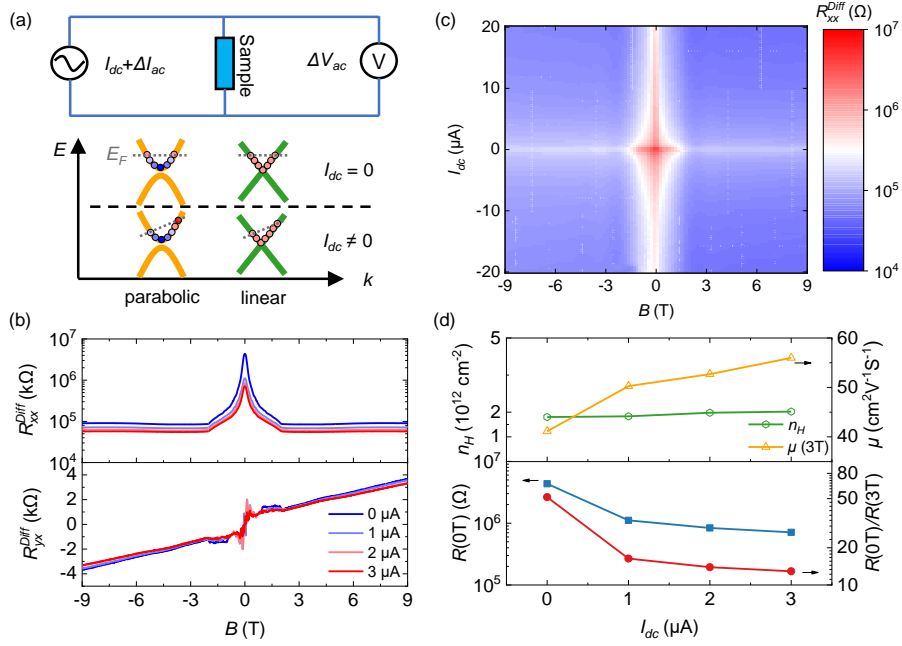

**FIG. 3. Transport properties of EuCd$_2$As$_2$ tuned by *dc* bias current $I_{dc}$ at 2 K.** (a) The upper panel is a schematic diagram of differential resistance measurement, and the lower panel is the comparison of parabolic dispersion and linear dispersion band. (b) Field sweeps of differential resistance $R_{xx}^{\text{Diff}}$ (upper panel) and $R_{yx}^{\text{Diff}}$ (lower panel) with $I_{dc}$ between 0 μA to 3 μA. (c) 2D map of $R_{xx}^{\text{Diff}}$ versus $I_{dc}$ and $B$. (d) The upper panel are $n_H$ (green hexagon) and mobility at 3 T $\mu(3\text{T})$ (yellow triangle) tuned by $I_{dc}$. The lower panel are differential resistance $R(0\text{T})$ (blue square) and $R(0\text{T})/R(3\text{T})$ (red circle) tuned by $I_{dc}$.

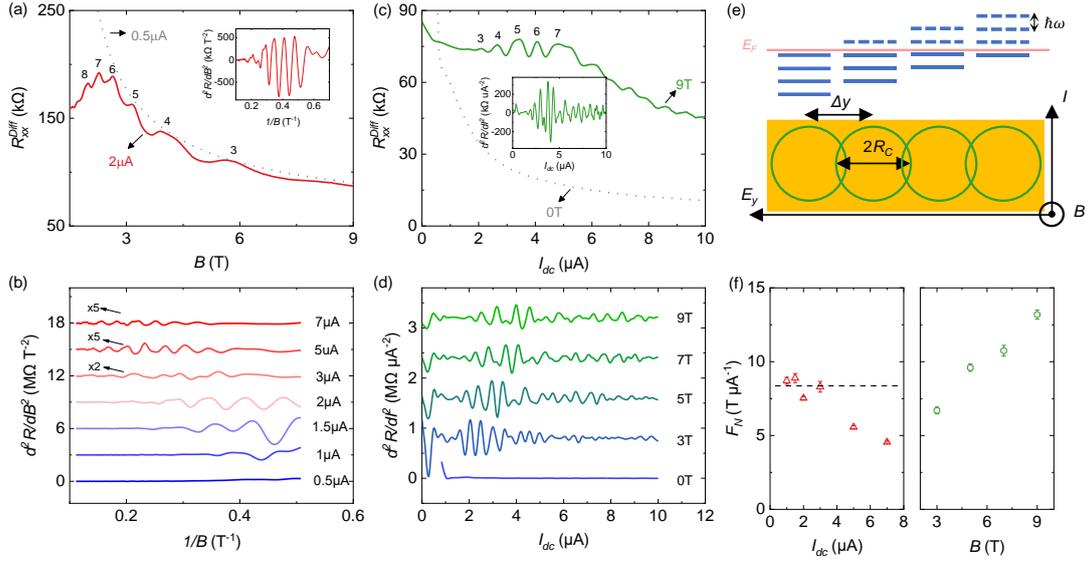

**FIG. 4. HIRO stimulated by *dc* bias current $I_{dc}$ in EuCd$_2$As$_2$ at 10K.** (a) Field sweep of differential resistance $R_{xx}^{Diff}$ with $I_{dc}$ 2 μA (red line) and 0.5 μA (gray dot line) (sample A3). The inset is the second derivative of $R_{xx}^{Diff}$ plotted as a function of *1/B*. (b) The second derivative of $R_{xx}^{Diff}$ with $I_{dc}$ between 0.5 μA and 7 μA. (c) *dc* current sweep of $R_{xx}^{Diff}$ with magnetic field 9 T (green line) and 0 T (gray dot line). The geometry ratio $\grave{o}_{dc}$ is labeled at each oscillation peak in (a) and (c). The inset is the second derivative of $R_{xx}^{Diff}$. (d) The second derivative of $R_{xx}^{Diff}$ with magnetic field between 0 T and 9 T. (e) Schematic diagram of geometric resonance of HIRO. Blue line, blue dashed line and pink line represent filled Landau level, unfilled Landau level and Fermi level, respectively. (f) Normalized frequency $F_N$ of HIRO with sweeping magnetic field (red triangle, left panel) and sweeping *dc* current (green hexagon, right panel). The dashed line is averaged $F_N$.